# Creating high density ensembles of nitrogen-vacancy centers in nitrogen-rich type Ib nanodiamonds

*Long-Jyun Su[1,5], Chia-Yi Fang[1,2,5], Yu-Tang Chang[1,2], Kuan-Ming Chen[3], Yueh-Chung Yu[3], Jui-Hung Hsu[4,6] and Huan-Cheng Chang[1,2,6]*

[1.] Institute of Atomic and Molecular Sciences, Academia Sinica, Taipei 106, Taiwan.

[2.] Department of Chemistry, National Taiwan University, Taipei 106, Taiwan.

[3.] Institute of Physics, Academia Sinica, Taipei 115, Taiwan.

[4.] Department of Materials and Optoelectronic Science, National Sun Yat-Sen University, Kaohsiung 804, Taiwan.

E-mail: jhsu@mail.nsysu.edu.tw; hchang@gate.sinica.edu.tw

**Abstract.** This work explores the possibility of increasing the density of negatively charged nitrogen-vacancy centers ([NV$^-$]) in nanodiamonds using nitrogen-rich type Ib diamond powders as the starting material. The nanodiamonds (10 − 100 nm in diameter) were prepared by ball-milling of microdiamonds, in which the density of neutral and atomically dispersed nitrogen atoms ([N$^0$]) was measured by diffuse reflectance infrared Fourier transform spectroscopy (DRIFTS). A systematic measurement for the fluorescence intensities and lifetimes of the crushed monocrystalline diamonds as a function of [N$^0$] indicated that the [NV$^-$] increases nearly linearly with [N$^0$] at 100 − 200 ppm. The trend, however, failed to continue for nanodiamonds with higher [N$^0$] (up to 390 ppm) but poorer crystallinity. We attribute the result to a combined effect of fluorescence quenching as well as the lower conversion efficiency of vacancies to NV$^-$ due to the presence of more impurities and defects in these as-grown diamond crystallites. The principles and practice of fabricating brighter and smaller fluorescent nanodiamonds (FNDs) are discussed.

---

[5] These two authors contribute equally to this work.
[6] Corresponding authors should be addressed.

1. **Introduction**

The negatively charged nitrogen-vacancy (NV⁻) defect in diamond is one of the best characterized color centers in the solid state [1-3]. It has found wide applications in areas such as single photon generation [4], quantum computing [5], nanoscale magnetic sensing [6, 7], and super-resolution far-field optical microscopy [8] recently. The center is also appealing for use in bioimaging owing to its unique characteristics of far-red emission and superb photostability [9]. Containing a high-density ensemble of NV⁻ centers, fluorescent nanodiamond (FND) is a nanomaterial developed for such application [10, 11]. The material, made of synthetic type Ib diamond, has excellent biocompatibility and exceptionally low toxicity [12, 13]. However, it has not yet received widespread attention in biology and biomedicine because its fluorescence intensity is relatively low compared with those of quantum dots, organic dyes, and fluorescent proteins [14, 15]. The main reason for this is that the absorption cross section of NV⁻ is about 1 − 2 orders of magnitude smaller than those of organic dyes (or fluorescent proteins) and quantum dots [16, 17]. A way to overcome this deficiency is to increase the density of NV⁻ centers, [NV⁻], in the particles. If each 10-nm FND can contain 10 NV⁻ centers, which is equivalent to [NV⁻] ~ 100 ppm, then the nanomaterial will become highly competitive in the field of bioimaging. However, such a goal is difficult to achieve because the typical density of atomically dispersed nitrogen in most commercially available type Ib diamonds synthesized by high pressure and high temperature (HPHT) methods is 100 ppm. In addition, the conversion efficiency of vacancies created by radiation damage to NV⁻ is low, typically in the range of 10% [18, 19]. The efficiency is expected to be even lower for nanoscale diamonds, where annihilation of vacancies can occur more readily at surface during annealing [20]. In view of this intrinsically limited conversion efficiency, we propose to use nanodiamond particles with a higher nitrogen density as the starting material to increase [NV⁻] [10, 21].

Infrared spectroscopy is a commonly used technique to characterize the density of nitrogen in diamond [22]. **Figure 1a** presents a typical infrared absorption spectrum of type Ib bulk diamond. Two prominent features are observed at 1130 and 1344 cm$^{-1}$ and they have been attributed to the localized vibrational modes of C-N bonds [23, 24]. The strengths of these two absorption bands are directly correlated with the density (in ppm) of atomically dispersed nitrogen in the neutral charge state as [22-25]

$$[N^0] = 37.5\mu_{1344} = 25\mu_{1130}, \tag{1}$$

where $\mu$ is the absorption coefficient (in cm$^{-1}$) of the sample at the specified wavenumber. This relation provides a foundation for the measurement of [N$^0$] in bulk diamond using infrared spectroscopy. However, it is inapplicable to submillimeter-sized diamond powders, whose sample thickness and hence the absorption coefficient is ill-defined. Liang et al. [26] have proposed an alternative based on the ratio of the two absorption coefficients at 1130 and 2120 cm$^{-1}$ as

$$[N^0] = 25 \times 5.5 \times \left(\frac{\mu_{1130}}{\Delta\mu_{2120}}\right), \tag{2}$$

where $\Delta\mu_{2120}$ is the absorption coefficient difference between the linear line drawn across the two peaks at 2033 and 2160 cm$^{-1}$ and the dip at 2120 cm$^{-1}$ (**Figure 1a**). The depth of this dip is characteristic of the two-phonon absorption band of diamond and therefore can be used as an internal standard. However, a complication in this analysis is that an infrared microscope coupled with a Fourier transform infrared (FTIR) spectrometer has to be used in order to acquire the IR absorption spectra of individual diamond microparticles, from which an ensemble average is obtained. The method is time-consuming and laborious and the smallest diamond that can be measured is ~150 μm in diameter.

To circumvent the aforementioned problems, we have developed a method based on diffuse reflectance infrared Fourier transform spectroscopy (DRIFTS), a technique widely used to collect and analyze IR photons scattered from fine particles and powders. The technique has been utilized to characterize the chemical functional groups on diamond surfaces as well [27, 28]. In a study of sonochemical reactions on diamond with DRIFTS, Visbal, et al. [28] observed the two-phonon absorption band as well as the C-N vibrations for synthetic type Ib diamond powders of grain size of 5 – 12 μm. For finer particles, the observation at the wavenumber of less than 2000 cm$^{-1}$ was obscured by the absorptions of surface functional groups (such as -C-O-C-, -C=O, etc.) and/or other adsorbed species after acid washes.  In this work, in order to minimize the interference of adsorbate with our measurement, we used particles with size greater than 5 μm and without any acid treatment. Moreover, we modified Eq. (2) to

$$[N^0] = 37 \times 5.5 \times \left( \frac{\Delta\mu_{1130}}{\Delta\mu_{2120}} \right), \qquad (3)$$

where $\Delta\mu_{1130}$ is the absorption coefficient difference corresponding to the height of the hump at 1130 cm$^{-1}$ (cf. **Figure 1a**).  The modification is justified based on the ratio of $\mu_{1130}/\mu_{1280} = 3.1$ for type Ib bulk diamond [23-26].  The advantage of this modified equation is that the measured $[N^0]$ is independent of the sample thickness and also immune to the baseline shift of the DRIFT spectra.

Based on the relationship, we present the result of our effort to fabricate brighter and smaller FNDs using N-rich type Ib diamond powders.  By carefully measuring (1) $[N^0]$ with DRIFTS, (2) fluorescence intensities with a home-built spectrometer, and (3) fluorescence lifetimes with a picosecond laser system, we confirm that the $[NV^-]$ in type Ib nanodiamonds (~100 nm in diameter) increases with increasing $[N^0]$ at 100 − 200 ppm.  However, the feature does not hold for as-grown diamonds with a density close to 400 ppm but poorer crystal quality.  The result is at variance with the recent report that an enormously high NV concentration of up to 1% can be attained for sintered detonation nanodiamonds with polycrystalline structure [29].  We furthermore developed a new method to crush nanodiamonds into smaller size in a gentle way. Our results indicate it an effective method to produce sub-20-nm nanodiamonds without loss of fluorescence intensity.

## 2. Experimental Section

*2.1. Materials*

Synthetic microdiamond powders were obtained from three different sources: Element Six (USA), Fine Abrasive Taiwan (Taiwan), and Changsha Naiqiang Superabrasives (China).  Five batches of diamonds with different grain sizes were tested and compared, including MDA, YK-J, PK-5, RVD1, and RVD2 (**Table 1**).  Additionally, pristine nanometer-sized diamond particles, Micron+MDA (0 − 0.1 μm) and MSY (0 – 0.05 μm) fabricated by Element Six (USA) and Microdiamant (Switzerland), respectively, were used as controls for comparison.

*2.2. Diffuse Reflectance Infrared Fourier Transform Spectroscopy (DRIFTS)*

IR spectra of microdiamond powders were acquired by using a FTIR spectrometer (MB-154, Bomem) equipped with a DRIFT cell (Spectra Tech) and a liquid-nitrogen-cooled HgCdTe (MCT) detector. Crystallites of KBr served as the sample to collect reference spectra.   A spectral resolution of 1 $cm^{-1}$ and a summation of 100 scans were adopted throughout the experiments.

*2.3. Ball milling*

Microdiamond powders were ground mechanically by using a high-energy shaker mill (8000M, SPEX) [30].   The grinding was conducted in a tungsten carbide (WC) vial containing 5 WC balls (10 mm in diameter).   The typical milling time was 30 min.

*2.4. Size separation*

NDs with size in the range of 100 nm were extracted from ball-milled microdiamonds by cleaning the particles in mixed *aqua regia*-hydrogen peroxide solution (1:19, v/v) heated gently to 85 °C for 1 h [31], followed by centrifugal fractionation at 4000 rpm.   From repeated sonication and centrifugal separation, a white colloidal solution containing well dispersed ND particles was obtained.   Size distributions of the particles were determined by dynamic light scattering using a combined particle size and zeta potential analyzer (Delsa Nano C, Beckman-Coulter).

*2.5. Transmission electron microscopy (TEM)*

NDs were transferred to the TEM grid by applying the particle suspension to the grid surface.   After being dried in air, they were characterized by using a field emission TEM (Tecnai F20 G2, FEI) operated at 200 keV and equipped with a 2k × 2k CCD camera (UltraScan 1000, Gatan).

*2.6. FND fabrication*

NDs were radiation-damaged with either a 3-MeV $H^+$ or a 40-keV $He^+$ ion beam, as detailed previously [32, 33]. In the first method, a thin diamond film (<50 μm thick) was prepared by depositing ~5 mg of the fine powders on a silicon wafer (1 × 1 $cm^2$) and subsequently subjected to the ion irradiation at a dose of ~2 × $10^{16}$ $H^+/cm^2$. In the second method, about 70 mg of the powders were deposited on a copper tape (2-m long and 35-mm wide) to form a thin film with an average thickness of ~0.2 μm. The tape rolled in vacuum, allowing the particles to be irradiated continuously by helium ions at a dose of ~2 × $10^{13}$ $He^+/cm^2$. Afterwards, the radiation-damaged diamonds were annealed at 800 °C for 2 h to form FNDs. To remove graphitic carbon atoms on surface, the freshly prepared FNDs were oxidized in air at 490 °C for 2 h and microwave-cleaned in concentrated $H_2SO_4$-$HNO_3$ (3:1, v/v) at 100 °C for 3 h.

*2.7. High-pressure crushing*

FNDs with size in the range of 30 nm were mixed with NaF powders at the weight ratio of 1:10 and pressed to form a pellet under a pressure of 10 tons by a hydraulic oil press (SPECAC). Next, the pellet was annealed at 720 °C in vacuum for 2 h, after which it was dissolved in hot water to remove NaF.

*2.8. Fluorescence spectroscopy*

Fluorescence spectra of FNDs suspended in water (typically 1 mg/mL) were acquired by using a setup built in-house [32, 33]. It consisted of two continuous-wave (cw) light sources, i.e. a 532-nm laser (DPGL-2100F, Photop Suwtech) and a 473-nm laser (LDC-1500, Photop Technologies), a set of dichroic beam splitters (Z532RDC, Chroma and FF484-FDi01-25x36, Semrock), a long-working distance microscope objective (50×, NA 0.55, Mitutoyo), a long-pass edge filter (E550LP, Chroma), and a multichannel analyzer (C7473, Hamamatsu). Backward fluorescence was collected through the same objective to avoid strong scattering to distort the spectra.

*2.9. Fluorescence lifetime measurement*

Fluorescence lifetimes of FNDs suspended in water were measured with a home-built microscope system and a 70-ps, 470-nm pulsed laser (LDH-P-C-470, PicoQuant) [34]. The system consisted of an inverted microscope (IX-71, Olympus) equipped with a dichroic filter (Z473RDC, Chroma) and a 4× objective (UPLSAPO, Olympus), which focused the laser light into the FND suspension in a shallow well mounted on a glass slide. Fluorescence emission was collected by the same objective and guided through two optical filters (HQ485LP, Chroma and NT62-987, Edmund Optics) onto an avalanche photodiode (SPCM-AQR-14-FC, PerkinElmer). Fluorescence lifetime data were recorded by using a photon counting card (SPC-830, Becker-Hickl) and analyzed with the SPC-Image software (Becker-Hickl).

## 3. Results and Discussion

### 3.1. Nitrogen density

**Figure 1b** presents the DRIFT spectrum of RVD1 microdiamonds (particle size of 210 − 250 μm) and its comparison with the direct IR absorption spectrum of type Ib bulk diamond with $[N^0]$ = 109 ppm [16]. In this comparison, the DRIFT spectrum was so scaled and shifted that the two-phonon absorption band of the microdiamonds overlaps well with that of the bulk. The difference in the intensity of the absorption bands at 1130 cm$^{-1}$ thus reflects the difference in $[N^0]$ between these two samples. In determining $[N^0]$ from the spectral comparison, caution has to be taken for the baseline shift of the DRIFT spectra as well as the interference from atmospheric water absorptions at 1400 − 1800 cm$^{-1}$. Use of Eq. (3) enables us to circumvent the problems and measure $[N^0]$ in these microdiamonds with an uncertainty of less than 20%.

**Table 1** summarizes the result of the DRIFTS measurements for five diamond samples from three different sources. The $[N^0]$ varies widely from 120 ppm of MDA to 386 ppm of PK-5. Moreover, the density varies up to 50 ppm between batches of the same product (e.g. RVD1 and RVD2 from the same manufacturer). As noted in the table, all samples except PK-5 were produced by crushing high-quality monocrystalline diamonds. PK-5, in contrast, consists of as-grown virgin diamond and therefore can contain a higher density of nitrogen and likely trap more metal impurities from catalysts used in the HPHT synthesis [35]. A study of the samples with transmission electron microscopy (TEM) revealed that both the MDA and PK-5 nanodiamonds (size ~ 100 nm) are monocrystalline in structure but irregular in shape (**Figure 2a** and **2b**). Some PK-5 particles do contain crystallographic defects such as the twin-like structure, as evidenced by the observation of the Moiré fringes in **Figure 2b** [35].

*3.2. Fluorescence intensity*

The microdiamonds after characterization with DRIFTS were crushed into finer grains by ball milling, separated by differential centrifugation, and converted into FNDs by ion irradiation and subsequent annealing. Dynamic light scattering measurement indicated that all particles are very similar in size, with a number-averaged diameter of ~100 nm (**Figure 3a**). **Figure 3b** compares the fluorescence spectra of one representative N-rich sample, RVD2, with that of Micron+MDA (0−0.1 μm), suspended in water and excited with a 532 nm laser. As noted, both the $NV^0$ and $NV^-$ centers are identifiable in the spectra based on their distinct zero phonon lines (ZPLs) at 575 nm and 638 nm, respectively [36, 37]. For the 100-nm FNDs prepared with crushed monocrystalline microdiamonds (such as MDA, YK-J, RVD1, and RVD2 listed in **Table 1**), we did observe a trend of the fluorescence intensity increase with increasing $[N^0]$ (**Figure 4a**). However, the trend failed to continue for samples, such as PK-5, with higher $[N^0]$ but poorer crystal quality. Depending on the irradiation treatment (either 40-keV $He^+$ or 3-MeV $H^+$), the fluorescence intensity of this sample is only comparable to or even lower than that of MDA (and also Micron+MDA), although its nitrogen density is more than 3 times as high.

The dose of the 3-MeV $H^+$ irradiation used in this experiment was $\sim 2 \times 10^{16}$ protons/cm$^2$, which resulted in an estimated damage density of $\sim 2 \times 10^{19}$ vacancies/cm$^3$ [16, 38]. Similar damage densities were achieved with the 40-keV $He^+$ irradiation [32, 33]. We have attempted to increase the dose by a factor of 2 but did not observe significant change in the fluorescence intensity. The result is in agreement with the report of Waldermann et al. [39] who used 2-MeV $He^+$ irradiation to create high density ensembles of $NV^-$ in type Ib bulk diamond and found that the optimum damage density for the formation of optically accessible $NV^-$ is $\sim 3 \times 10^{19}$ vacancies/cm$^3$. In **Figure 4a**, the reason there is a ~2-fold difference in fluorescence intensity between these two sets of data is perhaps due to the fact that the 3-MeV H$^+$ can penetrate much deeper in diamond than 40-keV He$^+$, i.e. 50 μm versus 0.20 μm [38], which makes it easier to ensure that all particles are irradiated. Moreover, the irradiation is much more gentle (~100× lower ion flux [32]), resulting in less amorphization of the diamond matrix. The latter can lead to fluorescence quenching and thus reduction of the fluorescence lifetime as discussed below.

*3.3. Fluorescence lifetime*

A possible cause for the marked decline of the fluorescence intensity at [N$^0$] = 386 ppm in **Figure 4a** is that the fluorescence quantum yield ($Q_F$) decreases with increasing [N$^0$]. The yield is related to the radiative and non-radiative decay rates, $k_r$ and $k_{nr}$ respectively, of the NV$^-$ center by

$$Q_F = \frac{k_r}{k_r + k_{nr}} = k_r \tau. \tag{4}$$

Since $k_r$ is a constant, measuring the fluorescence lifetime ($\tau$) thus provides a means to verify this possibility. Shown in **Figure 3c** are typical fluorescence decay time traces of two FND samples prepared by using either 40-keV He$^+$ or 3-MeV H$^+$ irradiation for the same MDA diamond. In this experiment, the fluorescence lifetimes were also measured for FNDs suspended in water to obtain ensemble averages. A picosecond laser operating at 470 nm and 5 MHz excited the NV$^-$ center, which has a phonon sideband extending to ~450 nm [36]. The resulting fluorescence was collected at the wavelength of $\lambda_{em}$ > 700 nm to avoid detection of the signal (<20%) from NV$^0$, which has a significantly longer emission lifetime than NV$^-$ (19 ns [40] versus 11.6 ns [41]) in synthetic type Ib bulk diamond. The observed fluorescence lifetime trace was then fitted with a double-exponential decay function, $I(t) = a_1\exp(-t/\tau_1) + a_2\exp(-t/\tau_2)$, from which the intensity-weighted mean lifetime ($\tau_i$) was calculated as

$$\tau_i = \frac{a_1\tau_1^2 + a_2\tau_2^2}{a_1\tau_1 + a_2\tau_2}. \tag{5}$$

**Figure 4b** shows the measured lifetimes of ten FND samples and their [N$^0$] dependence. For the 3-MeV $H^+$ irradiation, the longest lifetime observed for the four crushed monocrystalline nanodiamonds in water is $\tau_i \sim 17$ ns of MDA, which has the lowest $[N^0]$. The lifetime, however, substantially decreases to ~11 ns for PK-5, the as-grown virgin diamond. Similar results were obtained for FNDs prepared by using the 40-keV $He^+$ irradiation, although their corresponding lifetimes are all systematically shorter by ~20% [42].

The variation of the fluorescence lifetimes shown in **Figure 4b** should not be associated with the change in refractive index of the surrounding medium of the $NV^-$ centers [43, 44], since the FND particles used in these measurements are all similar in size (**Figure 3a**) and were examined under the same conditions. Rather, it is more likely to be a result of fluorescence quenching due to the presence of more impurities (such as nitrogen atoms and metal inclusions) and defects (such as twins, dislocations, etc.) in these as-grown diamond crystals (**Figure 2b**). It is noted that the observed lifetime shortening alone is not sufficient to account for the large intensity decrease of PK-5. If we assume that the measured fluorescence intensity ($I$) is linearly proportional to $[NV^-]$ and $Q_F$ as $I \propto [NV^-] \cdot Q_F$, then $[NV^-] \propto I/Q_F \propto I/\tau$, according to Eq. (4). With the measurements of both $I$ and $\tau$ for the individual samples, it is possible to determine the relative density of $NV^-$ among the FND particles. **Figure 4c** presents the results of the analysis, where the relative $[NV^-]$ is seen to increase nearly linearly with increasing $[N^0]$ at 100 – 200 ppm for crushed monocrystalline diamonds. In contrast, the $[NV^-]$ of PK-5 is much lower than expected, despite it containing up to 390 ppm of atomic nitrogen. It is speculated that the presence of structural defects in the as-grown diamond has a profound effect on the NV formation since they will reduce the conversion efficiency of $N^0 \to NV^-$ through vacancy annihilation at different types of the host-defect interfaces. Such an effect is anticipated to be more prominent for detonation nanodiamonds, which are polycrystalline in structure and contain a higher density of impurities [45, 46]. We conclude from the systematic intensity as well as lifetime measurements that it is the combination of the lower fluorescence quantum yield and the less efficient production of $NV^-$ that results in the weaker fluorescence as observed experimentally for PK-5 in **Figure 4a**.

*3.4. $NV^0$ versus $NV^-$*

For the formation of NV⁻ in radiation-damaged type Ib diamond, the widely accepted mechanism is that it involves a two-step reaction as [47, 48]

$$V^0 + 2N^0 \rightarrow NV^0 + N^0 \rightarrow NV^- + N^+ \qquad (6)$$

In the second reaction, an additional electron donor located in the vicinity of $NV^0$ is required to convert this defect into its negatively charged state, $NV^-$. The present study of the $[N^0]$ dependence provides a test ground for the theory. By comparing the fluorescence intensities of the bands at 575 nm (contributed mainly by the $NV^0$ centers) and 750 nm (contributed mainly by the $NV^-$ centers) in **Figure 3b**, we obtained an intensity ratio of $I_{575}/I_{750} \sim 0.4$ for the MDA diamond with $[N^0]$ = 120 ppm. The ratio, however, decreased noticeably to ~0.2 when $[N^0]$ increased to 386 ppm (**Figure 5a**), implying a decrease in relative density of $NV^0$ versus $NV^-$. To enhance the observation of this effect, a 473 nm laser was used to acquire the fluorescence spectra. The laser preferentially excites $NV^0$, thereby improving its fluorescence intensity (**Figure 5b**). Plotting $I_{575}/I_{750}$ against $[N^0]$ yields results in agreement with those of the 532-nm excitation for samples prepared with either $H^+$ or $He^+$ irradiation. Such a distinct $[N^0]$ dependence of $[NV^0]/[NV^-]$ supports the reaction mechanism illustrated in Eq. (6).

According to the above mechanism, the highest achievable $[NV^-]$ is 50 ppm for diamond containing 100 ppm of $N^0$. Additionally, the $[NV^-]$ should increase with $[N^0]$ if sufficient vacancies have been supplied by the ion irradiation. This leads to an interesting question: "*What is the highest possible density of NV⁻ in type Ib nanodiamond?*" In a study on the change of the absorption spectra of type Ib bulk diamond by heavy neutron irradiation, Mita [37] observed that the $[NV^-]$ increases linearly with the irradiation dose up to $7 \times 10^{17}$ neutrons/cm², and then decreases. At this optimum dose, the diamond with $[N^0]$ = 128 ppm showed an absorption band peaking at ~560 nm with an absorption coefficient of ~260 cm⁻¹ at 532 nm. This coefficient along with the absorption cross section of $\sigma_b = 3.1 \times 10^{-17}$ cm² [16] for $NV^-$ in bulk diamond yields $[NV^-]$ = 48 ppm. This density is much higher than 25 ppm estimated by Wee et al. [16], 6.3 ppm measured by Aharonovich et al. [49], and 13.5 ppm reported more recently by Botsoa et al. [50]. However, it should be taken as the upper limit of the density since complete conversion of the radiation-generated vacancies ($V^0$ and $V^-$) to $NV^-$ during thermal annealing had been assumed when estimating [NV⁻] by Davies et al. [51]. Nonetheless, the density is likely to be achieved by use of N-rich diamond.

*3.5. Size reduction*

Having both high [NV⁻] and high $Q_F$, the FNDs fabricated with RVD2 diamonds meet the requirements for practical use in bioimaging. If one makes a reasonable estimation for the conversion efficiency of $N^0 \rightarrow NV^-$ to be ~20%, the N-rich 100-nm diamonds can then contain up to 30 ppm of NV⁻ or ~ 3000 NV⁻ fluorophores per particle. Detecting these bright FND particles individually on a glass coverslip or even in cells is feasible with commercial confocal fluorescence microscopes [52]. For real-world applications as biolabels, the FNDs should be made as small as possible [53]. These particles can be either produced by crushing 100-nm FNDs into smaller parts with the ball-milling technique or made from smaller pristine NDs treated with the ion irradiation and annealing as described earlier. **Figure 6a** compares the fluorescence spectra of 30-nm and 35-nm FNDs, prepared by using these two different methods, with that of 100-nm FNDs as the reference sample. For the 30-nm FNDs prepared by ball milling, a 2-fold loss in fluorescence intensity (on the same weight basis) is resulted, which is in large part due to the decrease of the fluorescence lifetime (hence quantum yield) from $\tau_i = 17$ ns to $\tau_i = 13$ ns (**Figure 6b**). Compared with the 7-fold decrease in fluorescence intensity for FNDs prepared directly from pristine 35-nm nanodiamonds, the loss is clearly more acceptable.

It is tempting to fabricate smaller FNDs using the same ball-milling technique. Indeed, particles with diameters in the range of 15 nm can be produced, but their fluorescence intensity is ~10-fold lower than expected (on the same weight basis). This could be a result of surface-induced vacancy annihilation and crystal amorphization, both of which can occur more readily during ball milling of smaller FNDs. To overcome this hurdle, a method utilizing a hydraulic oil press to crush ball-milled 30-nm FNDs under a pressure of 10 tons was developed (described in section 2.7). The method is gentle and effective with a typical production yield of ~60% for sub-20-nm FNDs. Included in **Figure 6a** is the fluorescence measurement for 18-nm FNDs, whose fluorescence intensity and spectral profile are essentially unchanged after crushing. To further reduce the particle size to the sub-10 nm range without the loss of NV⁻, the oxygen etching techniques as described by Mohan et al. [15] and Gaebel et al. [54] could be used.

**4. Conclusion**

We have successfully fabricated FNDs containing high-density ensembles of NV⁻ centers in diamond nanoparticles of various size (10 − 100 nm) using N-rich type Ib diamond powders as the starting material. We demonstrate that it is possible to improve the brightness of FNDs with the [$N^0$] increasing from 100 ppm to 200 ppm for crushed monocrystalline diamonds. Our results indicate that through careful control of the radiation damage conditions and proper choice of the diamond materials, increasing [NV⁻] above 10 ppm in 10-nm FND particles is practical.

**Acknowledgement**

This work is supported by Academia Sinica and the National Science Council of Taiwan with Grant No. 99-2119-M-001-026- and 100-2112-M-110-002. We thank Ella Yang at Changsha Naiqiang Superabrasives for providing diamond samples and L.-C. Wang for taking the TEM images.

**References**

[1] Jelezko F and Wrachtrup J 2006 Single defect centers in diamond: A review  *Phys. Stat. Sol. (a)*  **203** 3207 - 3225.

[2] Aharonovich I, Castelletto S, Simpson DA, Su CH, Greentree AD and Prawer S 2011 Diamond-based single-photon emitters  *Rep. Prog. Phys.* **74** 076501.

[3] Aharonovich I, Greentree AD and Prawer S 2011 Diamond photonics  *Nature Photon.* **5** 397 - 405.

[4] Kurtsiefer C, Mayer S, Zarda P and Weinfurter H 2000 Stable solid-state source of single photons  *Phys. Rev. Lett.* **85** 290 - 293.

[5] Dutt MVG, Childress L, Jiang L, Togan E, Maze J, Jelezko F, Zibrov AS, Hemmer PR and Lukin MD 2007 Quantum register based on individual electronic and nuclear spin qubits in diamond  *Science* **316** 1312 - 1316.

[6] Balasubramanian G, Chan IY, Kolesov R, Al-Hmoud M, Tisler J, Shin C, Kim C, Wojcik A, Hemmer PR, Krueger A, Hanke T, Leitenstorfer A, Bratschitsch R, Jelezko F and Wrachtrup J 2008 Nanoscale imaging magnetometry with diamond spins under ambient conditions *Nature* **455** 648 - 651.

[7] Taylor JM, Cappellaro P, Childress L, Jiang L, Budker D, Hemmer PR, Yacoby A, Walsworth R and Lukin MD 2008 High-sensitivity diamond magnetometer with nanoscale resolution *Nature Phys.* **4** 810 - 816.

[8] Rittweger E, Han KY, Irvine SE, Eggeling C and Hell SW 2009 STED microsocpy reveals crystal colour centers with nanometric resolution *Nature Photon.* **3** 144 - 147.

[9] Hui YY, Cheng CL and Chang HC 2010 Nanodiamonds for optical bioimaging *J. Phys. D: Appl. Phys.* **43** 374021.

[10] Yu SJ, Kang MW, Chang HC, Chen KM and Yu YC 2005 Bright fluorescent nanodiamonds: no photobleaching and low cytotoxicity *J. Am. Chem. Soc.* **127** 17604 - 17605.

[11] Vaijayanthimala V and Chang HC 2009 Functionalized fluorescent nanodiamonds for biomedical applications *Nanomed.* **4** 47 - 55.

[12] Vaijayanthimala V, Tzeng YK, Chang HC and Li CL 2009 The biocompatibility of fluorescent nanodiamonds and their mechanism of cellular uptake *Nanotechnology* **20** 425103.

[13] Mohan N, Chen CS, Hsieh HH, Wu YC and Chang HC 2010 In vivo imaging and toxicity assessments of fluorescent nanodiamonds in *Caenorhabditis elegans* *Nano Lett.* **10** 3692 - 3699.

[14] Faklaris O, Garrot D, Joshi V, Boudou JP, Sauvage T, Curmi PA and Treussart F 2009 Comparison of the photoluminescence properties of semiconducotr quantum dots and non-blinking diamond nanoparticles. Observation of the diffusion of diamond nanoparticles in living cells *J. Eur. Opt. Soc. Rapid Pub.* **4** 09035.

[15] Mohan N, Tzeng YK, Yang L, Chen YY, Hui YY, Fang CY and Chang HC 2010 Sub-20-nm fluorescent nanodiamonds as photostable biolabels and fluorescence resonance energy transfer donors *Adv. Mater.* **22** 843 - 847.

[16] Wee TL, Tzeng YK, Han CC, Chang HC, Fann W, Hsu JH, Chen KM and Yu YC 2007

Two-photon excited fluorescence of nitrogen-vacancy centers in proton-irradiated type Ib diamond    *J. Phys. Chem. A* **111** 9379 - 9386.

[17] Chapman R and Plakhotnik T 2011 Quantitative luminescence microscopy on nitrogen-vacancy centers in diamond: saturation effects under pulsed excitation    *Chem. Phys. Lett.* **507** 190 - 194.

[18] Acosta VM, Bauch E, Ledbetter MP, Santori C, Fu KMC, Barclay PE, Beausoleil RG, Linget H, Roch JF, Treussart F, Chemerisov S, Gawlik W and Budker D 2009 Diamonds with a high density of nitrogen-vacancy centers for magnetometry applications    *Phys. Rev. B* **80** 115202.

[19] Pezzagna S, Naydenov B, Jelezko F, Wrachtrup J and Meijer J 2010 Creation efficiency of nitrogen-vacancy centres in diamond    *New J. Phys.* **12** 065017.

[20] Smith BR, Inglis DW, Sandnes B, Rabeau JR, Zvyagin AV, Gruber D, Noble CJ, Vogel R, Osawa E and Plakhotnik T 2009 Five-nanometer diamond with luminescent nitrogen-vacancy defect centers    *Small* **5** 1649 - 1653.

[21] Boudou JP, Curmi PA, Jelezko F, Wrachtrup J, Aubert P, Sennour M, Balasubramanian G, Reuter R, Thorel A and Gaffet E 2009 High yield fabrication of fluorescent nanodiamonds    *Nanotechnology* **20** 235602.

[22] Davies G 1999 Current problems in diamond: towards a quantative understanding    *Physica B* **273-274** 15 - 23.

[23] Woods GS, van Wyk JA and Collins AT 1990 The nitrogen-content of type-1b synthetic diamond    *Phil. Mag. B* **62** 589 - 595.

[24] Kiflawi I, Mayer AE, Spear PM, van Wyk JA and Woods GS 1994 Infrared-absorption by the signel nitrogen and a defect centers in diamond    *Phil. Mag. B* **69** 1141 - 1147.

[25] Lawson SC, Fisher D, Hunt DC and Newton M 1998 On the existence of positively charged single-substitutional nitrogen in diamond    *J. Phys.: Condens. Matter* **10** 6171 - 6180.

[26] Liang ZZ, Jia X, Ma HA, Zang CY, Zhu PW, Guan QF and Kanda H 2005 Synthesis of HPHT diamond containing high concentrations of nitrogen impurities using $NaN_3$ as dopant in metal-carbon system    *Diamond Relat. Mater.* **14** 1932 - 1935.

[27] Ando T, Inoue S, Ishii M, M. Kamo M, Sato Y, Yamada O and Nakano T 1993 Fourier-transform infrared photoacoustic studies of hydrogenated diamond surfaces    *J. Chem.*

*Soc., Faraday Trans.* **89** 749 - 751.

[28] Visbal H, Ishizaki C and Ishizaki K 2004 Ultrasonic treatment of acid-washed diamond powder surface   *J. Ceramic Soc. Japan* **112** 95 - 98.

[29] Baranov PG, Soltamova AA, Tolmachev DO, Romanov NG, R. Babunts A, Shakhov FM, Kidalov SV, Vul' AY, Mamin GV, Orlinskii SB and Silkin NI 2011 Enormously high concentrations of fluorescent nitrogen-vacancy centers fabricated by sintering of detonation nanodiamonds   *Small* **7** 1533 - 1537.

[30] Silva KL, Bernardi LO, Yokoyama M, Trombini V, Cairo CA and Pallone EMDA 2008 Obtained of diamond nanometric powders using high energy milling for the production of alumina-diamond nanocomposites   *Materials Science Forum* **591-593** 766 - 770.

[31] Archer M, McCrindle RI and Rohwer ER 2003 Analysis of cobalt, tanatlum, titaniun, vanadium and chromium in tungsten carbide by inductively coupled plasma-optical emission spectrosocpy *J. Anal. Atom. Spectrom.* **18** 1493 - 1496.

[32] Chang YR, Lee HY, Chen K, Chang CC, Tsai DS, Fu CC, Lim TS, Tzeng YK, Fang CY, Han CC, Chang HC and Fann W 2008 Mass produciton and dynamic imaging of fluorescent nanodiamonds   *Nature Nanotech.* **3** 284 - 288.

[33] Wee TL, Mau YW, Fang CY, Hsu HL, Han CC and Chang HC 2009 Preparation and characterization of green fluorescent nanodiamonds for biological applications   *Diamond Relat. Mater.* **18** 567 - 573.

[34] Hsu JH, Su WD, Yang KL, Tzeng YK and Chang HC 2011 Nonblinking green emission from single H3 color centers in nanodiamonds   *Appl. Phys. Lett.* **98** 193116.

[35] Yin LW, Li MS, Cui JJ, Bai YJ, Xu B, Gong JH and Hao ZY 2002 Planar defects and dislocations in HPHT as-grown diamond crystals   *Diamond Relat. Mater.* **11** 268 - 272.

[36] Davies G and Hamer MF 1976 Optical studies of 1.945 eV vibronic band in diamond   *Proc. R. Soc. A* **348** 285 - 298.

[37] Mita Y 1996 Change of absorption spectra in type-1b diamond with heavy neutron irradiation *Phys. Rev. B* **53** 11360 - 11364.

[38] http://www.srim.org/

[39] Waldermann FC, Olivero P, Nunn J, Surmacz K, Wang ZY, Jaksch D, Taylor RA, Walmsley IA, Draganski M, Reichart P, Greentree AD, Jamieson DN and Prawer S 2007 Creating diamond color centers for quantum optical applications   *Diamond Relat. Mater.* **16** 1887 - 1895.

[40] Liaugaudas G, Davies G, Suhling K, Khan RUA and Evans DJF 2012 Luminescence lifetimes of neutral nitrogen-vacancy centres in synthetic diamond containing nitrogen   *J. Phys.: Condens. Matter* **24** 435503.

[41] Collins AT, Thomaz MF and Jorge MIB 1983 Luminescence decay time of the 1.945 eV center in type 1b diamond   *J. Phys. C: Solid State* **16** 2177 - 2181.

[42] The decrease of the fluorescence lifetime with increasing $[N^0]$ is similarly found for amplitude-weighted mean lifetimes defined as $\tau_m = (a_1\tau_1 + a_2\tau_2)/(a_1 + a_2)$.

[43] Beveratos A, Brouri R, Gacoin T, Poizat JP and Grangier P 2001 Nonclassical radiation from diamond nanocrystals   *Phys. Rev. A* **64** 061802.

[44] Tisler J, Balasubramanian G, Naydenov B, Kolesov R, Grotz B, Reuter R, Boudou JP, Curmi PA, Sennour M, Thorel A, Börsch M, Aulenbacher K, Erdmann R, Hemmer PR, Jelezko F and Wrachtrup J 2010 Fluorescence and spin properties of defects in single digit nanodiamonds   *ACS Nano* **3** 1959 - 1965.

[45] Vlasov II, Shenderova O, Turner S, Lebedev OI, Basov AA, Sildos I, Rähn M, Shiryaev AA and Van Tendeloo G 2010 Nitrogen and luminescent nitrogen-vacancy defects in detonation nanodiamond   *Small* **6** 687 - 694.

[46] Bradac C, Gaebel T, Naidoo N, Sellars MJ, Twamley J, Brown LJ, Barnard AS, Plakhotnik T, Zvyagin AV and Rabeau JR 2010 Observation and control of blinking notrogen-vacancy centres in discrete nanodiamonds   *Nature Nanotech.* **5** 345 - 349.

[47] Collins AT 2002 The Fermi level in diamond   *J. Phys.: Condens. Matter* **14** 3743 - 3750.

[48] Collins AT, Connor A, Ly CH, Shareef A and Spear PM 2005 High-temperature annealing of optical centers in type-1 diamond   *J. Appl. Phys.* **97** 083517.

[49] Aharonovich I, Santori C, Fairchild BA, Orwa J, Ganesan K, Fu KMC, Beausoleil RG, Greentree AD and Prawer S 2009 Producing optimized ensembles of nitrogen-vacancy color centers for quantum information applications   *J. Appl. Phys.* **106** 124904.

[50] Botsoa J, Sauvage T, Adam MP, Desgardin P, Leoni E, Courtois B, Treussart F and Barthe MF 2011 Optimal conditions for NV- center formation in type-1b diamond studied using photoluminescence and positron annihilation spectroscopies   *Phys. Rev. B* **84** 125209.

[51] Davies G, Lawson SC, Collins AT, Mainwood A and Sharp SJ 1992 Vacancy-related centers in diamond   *Phys. Rev. B* **46** 13157 - 13170.

[52] Zhang B, Li Y, Fang CY, Chang CC, Chen CS, Chen YY and Chang HC 2009 Receptor-mediated cellular uptake of folate-conjugated fluorescent nanodiamonds: a combined ensemble and single-particle study   *Small* **5** 2716 - 2721.

[53] Fu CC, Lee HY, Chen K, Lim TS, Wu HY, Lin PK, Wei PK, Tsao PH, Chang HC and Fann W 2007 Characterization and application of single fluorescent nanodiamonds as cellcular biomarkers   *Proc. Natl. Acad. Sci. USA* **104** 727 - 732.

[54] Gaebel T, Bradac C, Chen J, Say JM, Brown L, Hemmer P and Rabeau JR 2011 Size-reduction of nanodiamonds via air oxidation   *Diamond Relat. Mater.* **21** 28 - 32.

**Table 1.** Properties of type Ib diamonds and their nitrogen densities measured by DRIFTS

| Names | Manufacturers | Size (μm) | $[N^0]$ (ppm) | Remarks |
|---|---|---|---|---|
| Micron+ MDA | Element Six | 0 − 0.1 | U.D.[7] | Crushed monocrystalline diamond |
| MDA | Element Six | 8 − 16 | 120 | Crushed monocrystalline diamond |
| YK-J | Fine Abrasive Taiwan | 20 − 30 | 151 | Crushed monocrystalline diamond |
| RVD2 | Changsha Naiqiang Superabrasives | 210 − 250 | 157 | Crushed monocrystalline diamond |
| RVD1 | Changsha Naiqiang Superabrasives | 210 − 250 | 209 | Crushed monocrystalline diamond |
| PK-5 | Fine Abrasive Taiwan | 20 − 30 | 386 | As-grown virgin diamond |

---

[7] Undetermined.

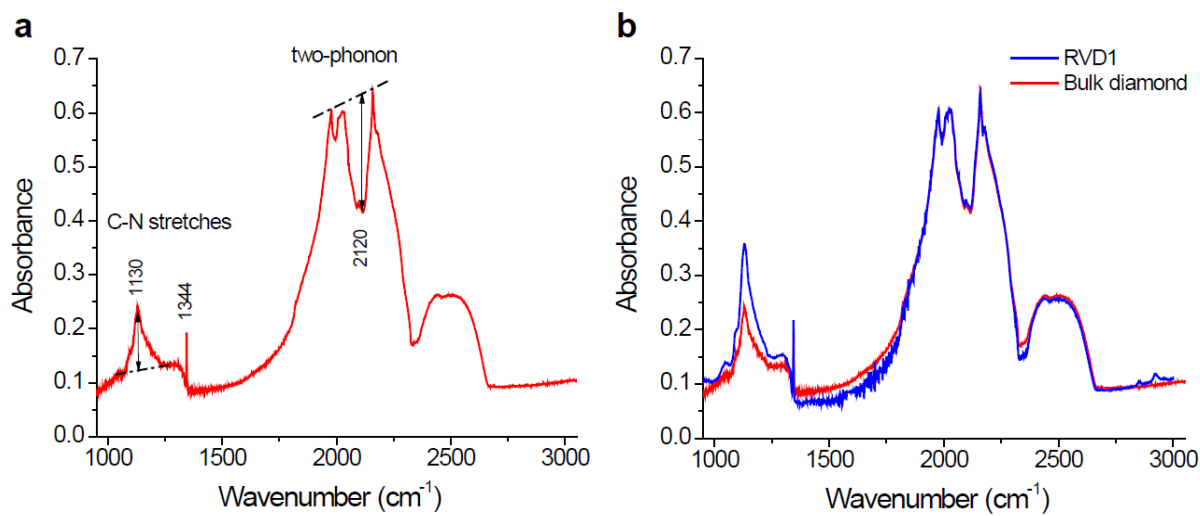

**Figure 1.** (a) IR absorption spectrum of a bulk diamond with $[N^0]$ = 109 ppm, determined by using Eq. (1) in text. The ratio of the depth of the dip at 2120 cm$^{-1}$ versus the height of the hump at 1130 cm$^{-1}$ was measured and used as a reference for the estimation of the density of substitutional nitrogen atoms in microdiamonds by DRIFTS. (b) DRIFT spectrum (blue) of a typical microdiamond sample (particle size of 210 − 250 μm) and its comparison with the absorption spectrum (red) of the bulk diamond in (a). The sharp features appearing at 1400 − 1800 cm$^{-1}$ and ~2350 cm$^{-1}$ are due to incomplete cancelation of atmospheric $H_2O$ and $CO_2$ absorption lines, respectively, between sample and reference spectra.

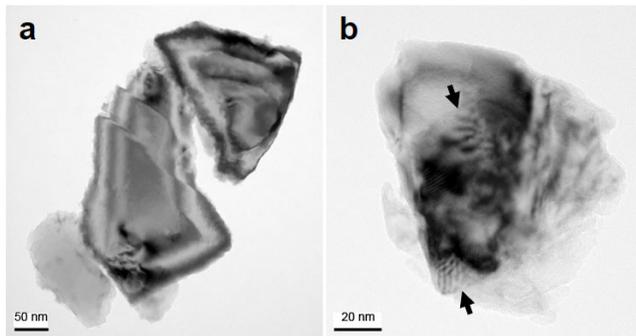

**Figure 2.** TEM images of nanodiamonds fabricated by ball-milling of microdiamonds: (a) MDA and (b) PK-5. The image in (a) shows that both the MDA particles under examination are monocrystalline with almost no structural defects. In comparison, the Moiré fringes (denoted by black arrows) observed for PK-5 diamond in (b) indicate the presence of twin-like structure.

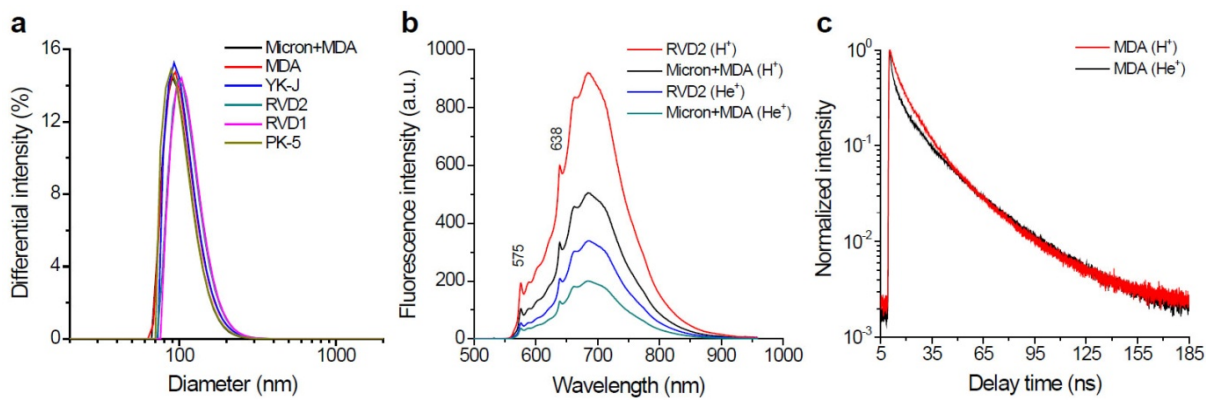

**Figure 3.** (a) Size distributions of FND particles produced by ball milling of type Ib microdiamonds listed in **Table 1** as the starting materials. (b) Comparison of the fluorescence spectra of four representative 100-nm FND samples suspended in water (1 mg/mL). The excitation was made with a cw 532 nm laser and the samples were prepared under two different irradiation conditions with either 3-MeV $H^+$ or 40-keV $He^+$ as the damage agent. (c) Fluorescence decay time traces of two representative 100-nm FND samples suspended in water (1 mg/mL). The excitation was made with a 470-nm picosecond laser operating at 5 MHz and the resulting fluorescence was collected at $\lambda_{em} > 700$ nm.

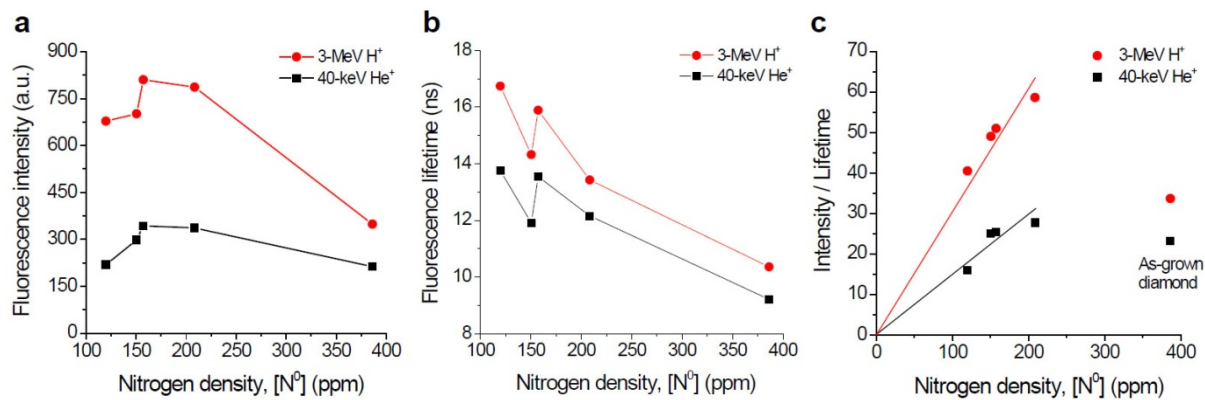

**Figure 4.** Fluorescence intensity (a) and lifetime (b) of 100-nm FNDs as a function of $[N^0]$. All the measurements were made for FNDs suspended in water at a concentration of 1 mg/mL. (c) Plots of fluorescence intensity/lifetime against $[N^0]$ in 100-nm FNDs. The intensity/lifetime ratio is linearly proportional to $[NV^-]$ as discussed in text. Note that the FNDs prepared by using either 3-MeV $H^+$ or 40-keV $He^+$ irradiation show similar $[N^0]$ dependence.

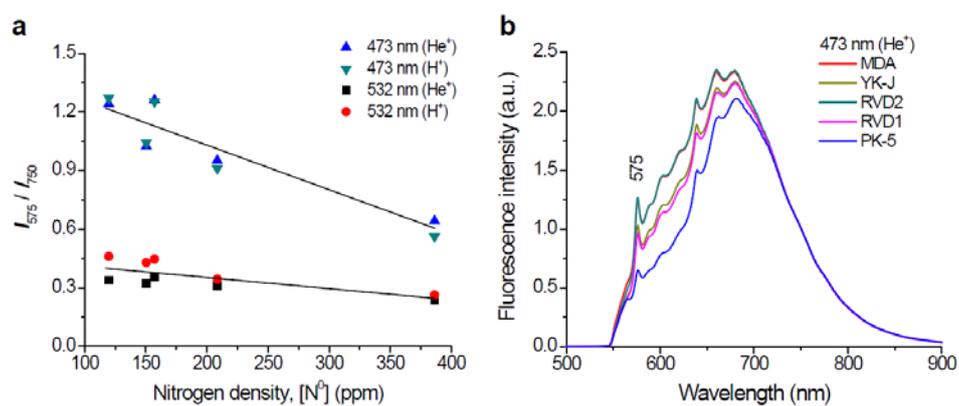

**Figure 5.** (a) Fluorescence intensity ratio of $NV^0$ (at 575 nm) versus $NV^-$ (at 750 nm) as a function of $[N^0]$. (b) Fluorescence spectra of 100-nm FNDs suspended in water (1 mg/mL), normalized at 750 nm. The FNDs of five samples were all prepared by 40-keV $He^+$ irradiation and then photoexcited at 473 nm to obtain the fluorescence spectra.

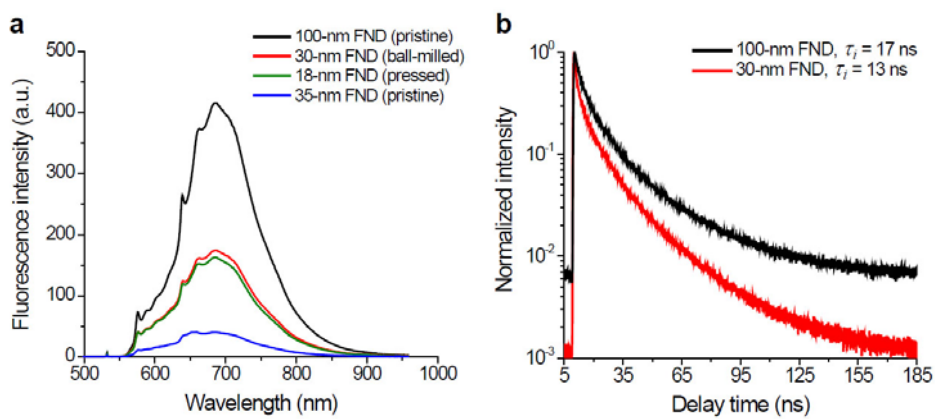

**Figure 6.** Comparison of the fluorescence intensities (a) and lifetimes (b) of FNDs with different sizes. The 100-nm and 35-nm FNDs were prepared from pristine NDs, 30-nm FNDs were produced from ball-milling of 100-nm FNDs, and 18-nm FNDs were fabricated by crushing of 30-nm FNDs with a hydraulic press, respectively. The concentration of all FND suspensions is 1 mg/mL.